\RequirePackage{fix-cm}
\documentclass[natbib, smallextended]{svjour3} 

\usepackage{amsmath,amssymb,amsfonts}
\usepackage{algorithmic}
\usepackage{graphicx}
\usepackage{textcomp}
\usepackage{multirow}
\usepackage{xcolor}
\usepackage[utf8]{inputenc}
\usepackage{lscape} 
\usepackage{color}

\smartqed  

\usepackage[most]{tcolorbox}
\usepackage{lineno}

\newcommand*{\affaddr}[1]{#1} 
\newcommand*{\affmark}[1][*]{\textsuperscript{#1}}

\usepackage[colorlinks = true,
            linkcolor = blue,
            urlcolor  = blue,
            citecolor = blue,
            anchorcolor = blue]{hyperref}

\begin{document}

\title{Robust Acoustic Domain Identification with its Application to Speaker Diarization}

\author{A Kishore Kumar\affmark[1] \and Shefali Waldekar\affmark[2] \and Md Sahidullah\affmark[3] \and Goutam Saha\affmark[1]}

\authorrunning{A Kishore Kumar \and Shefali Waldekar \and Goutam Saha \and
        Md Sahidullah 
}

\institute{
\at
\affaddr{\affmark[1] Department of Electronics \& ECE, Indian Institute of Technology Kharagpur, India.\\
           \email{kishore@iitkgp.ac.in, gsaha@ece.iitkgp.ac.in}}\\
\affaddr{\affmark[2] Department of Electrical, Electronics \& Communication Engineering, GITAM School of Technology, GITAM (Deemed to be University), Bengaluru, India.\\
           \email{swaldeka@gitam.edu}}\\
\affaddr{\affmark[4] Universit\'{e} de Lorraine, CNRS, Inria, LORIA, F-54000, Nancy, France.\\
           \email{md.sahidullah@inria.fr}}\\
}

\maketitle

\begin{abstract}
With the rise in multimedia content over the years, more variety is observed in the recording environments of audio. An audio processing system might benefit when it has a module to identify the acoustic domain at its front-end. In this paper, we demonstrate the idea of \emph{acoustic domain identification} (ADI) for \emph{speaker diarization}. For this, we first present a detailed study of the various domains of the third DIHARD challenge highlighting the factors that differentiated them from each other. Our main contribution is to develop a simple and efficient solution for ADI. In the present work, we explore speaker embeddings for this task. Next, we integrate the ADI module with the speaker diarization framework of the DIHARD III challenge. The performance substantially improved over that of the baseline when the thresholds for agglomerative hierarchical clustering were optimized according to the respective domains. We achieved a relative improvement of more than $5\%$ and $8\%$ in DER for core and full conditions, respectively, on Track 1 of the DIHARD III evaluation set.
\end{abstract}


\section{Introduction}
Spoken documents refer to audio recordings of multi-speaker events like lectures, meetings, news, etc. We use \emph{acoustic domain} as a descriptor for the type of spoken document. Although it sounds similar to \emph{acoustic scene}, it is not so because the latter concerns only the background environment of the audio which may or may not have speech~\citep{barchiesi2015acoustic}. In spoken documents, speech is the main source of information while other elements such as location and channel form the metadata. Examples of acoustic domains are broadcast news, telephone conversations, web video, movies, and so on, which may have different backgrounds across the data from same class, or even in the same document. Nowadays, easily accessible and simple to use audio recorders have led to an exponential increase in the amount of speech recordings which are available in myriad conditions in terms of environment, recording equipment, number of speakers, to name a few. An artificial intelligence system can deal with the multi-domain scenarios either by normalizing to the domains, i.e. \textit{domain-invariant processing}, or by adapting to the domains i.e. \textit{domain-dependent processing}. For diverse and challenging recording conditions, it is difficult to design a system following the first strategy. Therefore, in such cases, a system that identifies the domain and then proceeds with relevant information extraction accordingly is expected to perform better than a \emph{one-size-fits-all} method.

For \emph{rich transcription} (RT) of spoken documents, besides ``what is being spoken" and ``by whom", it is also imperative to know ``who spoke when". The answers to the three questions are provided by automatic speech recognition, speaker recognition, and speaker diarization, respectively. \emph{Speaker diarization} is the task of generating time-stamps with respect to the speaker labels~\citep{Anguera2012, PARK2022101317}. In earlier studies, RT researchers mainly considered recordings of broadcast news, telephone conversations and meetings/conferences~\citep{Anguera2012}. The NIST RT challenges (RT02 - RT09)\footnote{\url{https://www.nist.gov/itl/iad/mig/rich-transcription-evaluation}} were organized for metadata extraction from these acoustic domains. There, the type of data was known beforehand which implies that the diarization systems were designed to suit the application. The factors mainly contributing to poor diarization were number of speakers, speech activity detection, amount of overlap, and speaker modelling \citep{mirghafori2006nuts, huijbregts2007blame, sinclair2013challenges}. 

In this paper, we propose a simple but efficient method for acoustic domain identification (ADI). The proposed ADI system uses speaker embeddings. Our study reveals that i-vector and OpenL3 embedding based method achieves considerably better performance than x-vector based approach in the third DIHARD challenge dataset. Of all the three, i-vector system's results were found to be better. Next, we show that domain-dependent threshold for speaker clustering helps to improve the diarization performance only when \emph{probabilistic linear discriminant analysis} (PLDA) adaptation uses audio-data from all the domains. The proposed system stands apart from other submissions to the challenge because unlike most of the top-performers, it is not an ensemble or fusion-based system. Moreover, from the study of speaker diarization literature, we found that the domain-dependent processing approach is the first of its kind \citep{sahidullah2019speed}.

In the upcoming sections, first, we talk about previous studies relevant to the topic, i.e. acoustic scene classification and speaker diarization in Section~\ref{sec:lit}. Then, Section~\ref{sec:database} presents a detailed description of the speaker diarization data used in our study. The major aspects of the contribution of this work, that is, ADI and classification of domains are discussed in Section~\ref{sec:adi} and Section~\ref{sec:classify} respectively. The experimental setup is outlined in Section~\ref{sec:expt} while the results of the experiments are presented and discussed in Section~\ref{sec:result}. We conclude the paper in Section~\ref{sec:conc}. 

\section{Related works}
\label{sec:lit}
In the classical audio processing applications, where speech is regarded as the primary information source, the domain was kept constant to get the best performance. However, in the present times of technological advancements, multi-domain data is unavoidable, whether it is for speech recognition, language recognition, speaker recognition, or speaker diarization. In this paper, we consider the speaker diarization (SD) application. The research in this area took a leap two decades ago with the NIST RT evaluations. Back then, broadcast news (BN), conversational telephone speech (CTS), and meetings/conferences were the only three acoustic domains considered. 

The use of \emph{total variability} (TV) space was proposed in \citep{shum2011exploiting} to exploit intra-conversation variability. The same authors later applied Bayesian-\emph{Gaussian mixture model} (GMM) to \emph{principal component analysis} (PCA)-processed i-vectors as a probabilistic approach to speaker clustering \citep{shum2013unsupervised}. The authors of \citep{sell2014speaker} incorporated PLDA for i-vector scoring and used unsupervised calibration of the PLDA scores to determine the clustering stopping criterion. They also showed improved diarization by denser sampling in the i-vector space with overlapping temporal segments. A \emph{deep neural network} (DNN) architecture was introduced by \citep{garcia2017speaker} that learned a fixed-dimensional embedding for variable length acoustic segments along with a scoring function for measuring the likelihood that the segments originated from the same or different speakers. Combination of long short-term memory based speaker-discriminative d-vector embeddings and non-parametric clustering was done by \citep{wang2018speaker} while unbounded interleaved-state \emph{recurrent neural network} (RNN) with d-vectors were employed by \citep{zhang2019fully} for SD.

The annual DIHARD challenges focusing on \emph{hard diarization} revamped SD research. The objective is to build SD systems for spoken documents from diverse and challenging domains where the existing state-of-the-art systems fail to deliver. The first one was held in 2018 \citep{ryant2018first}. The data for this challenge was taken from nine domains. The top performing system \citep{sell2018diarization} explored the i-vector and x-vector representations and found that the latter trained on wideband microphone data was better. They also showed that a \emph{speech activity detection} (SAD) system specifically trained for SD with in-domain data helped in improving performance. For the second challenge, the data was drawn from 11 domains consisting of single-channel and multi-channel audio recordings \citep{ryant2019second}. This version was provided with baseline systems for speech enhancement, SAD and SD. The given SD baseline was based on the top-performing system of its predecessor. The first rank system performed \emph{agglomerative hierarchical clustering} (AHC) over x-vectors with Bayesian hidden Markov model  \citep{landini2019but}. In the third challenge, the multichannel audio condition and also the child speech data was not there, whereas CTS data was included for the first time \citep{ryant2020third}. The top performing system had speech enhancement and audio domain classification before SD \citep{wang2021ustc}. It combined multiple front-end techniques, such as speech separation, target-speaker based voice activity detection and iterative data purification. Its SD system was based on DIHARD II's top-ranked system.

The challenges on detection and classification of acoustic scenes and events (DCASE) have rendered a boost to research in \emph{acoustic scene classification} (ASC). Discussion about the state-of-the-art in this field is necessary before we go into the details of ADI because the two bear some degree of conceptual similarity. The baseline systems provided with the ASC task of the DCASE challenges have ranged from \emph{Mel-frequency cepstral coefficients} (MFCC)-GMM based systems \citep{giannoulis2013database, mesaros2016tut}, log mel-band energy with \emph{mulltilayer perceptron} (MLP) \citep{mesaros2018acoustic}, with convolutional neural network (CNN) \citep{mesaros2018multi, mesaros2019acoustic} to OpenL3 embeddings with two fully-connected feed-forward layers \citep{Heittola2020}. On the other hand, the top-performing systems have shown a variety of approaches to extract as much as possible differences between the given acoustic scenes. These approaches include the use of spectrograms, \emph{recurrence quantification analysis} (RQA) of MFCC, i-vectors, log-mel band energies, and constant-Q transform as features, with support vector machine, MLP, DNN, CNN, RNN, and trident residual neural network as classifiers \citep{Roma2013, Eghbal-Zadeh2016, Mun2017, Sakashita2018, Chen2019, Suh2020}. A few systems have also used data augmentation techniques like generative adversarial network, graph neural network, temporal cropping and mixup \citep{Mun2017, Sakashita2018, Chen2019, Suh2020}. The multi-classifier systems have employed majority vote, average vote, or random forest for decision making \citep{Mun2017, Sakashita2018, Chen2019, Suh2020}. Some other noteworthy experiments have been done with various spectral and time-frequency features, classification techniques along with fusion approaches for ASC \citep{ren2018deep, rakotomamonjy2015histogram, waldekar2018classification, waldekar2020analysis}. Matrix factorization was applied in \citep{bisot2017feature}. i-vector and x-vector embeddings were used with CNN in \citep{dorfer2018acoustic} and \citep{zeinali2018convolutional} respectively.

\section{DIHARD III: A large-scale realistic speech corpora with multiple domains}
\label{sec:database}
In the third edition of the DIHARD challenge series, the DIHARD III data set is taken from eleven diverse domains. This brings a lot of variability in data conditions like background noise, language, source, number of speakers, amount of overlapped speech, and speaker demographics. Table~\ref{Table:dtdesc} enlists the acoustic domains along with their corresponding source corpus, type of recording, and a brief description. The data for `Meeting' and `Sociolinguistic (Field)' are taken from two different speech corpora. The table also shows how the domains differ in terms of the number of speakers per sample and total speakers. The database is divided into development and evaluation sets. Both have 5--10 minute duration samples, except in the case of `Web videos’ where the sample duration ranges from less than 1 min to more than 10 min. Evaluation has to be done on two partitions of the evaluation data: Core evaluation set --- a balanced set where the total duration of each domain is approximately equal, and Full evaluation set --- all the samples are considered for each domain. The development set composition is the same as that of the evaluation set. The metadata provided for each data sample in the development set is its domain, source, language, and whether or not it was selected for the core set. This information was not provided for the evaluation data.

\section{Audio domain analysis of the corpora}
\label{sec:adi}

The heterogeneity in the recording conditions of the audio samples under consideration can be observed from the last two columns of Table~\ref{Table:dtdesc}. To gain better insights into their domain-wise differences, we analyzed the DIHARD III development data's audio samples using three methods: \emph{signal-to-noise ratio} (SNR) estimation, \emph{long-term average spectrum} (LTAS) analysis, and speech-to-non-speech ratio computation.

 \begin{landscape}
\thispagestyle{plain}
\begin{table}[t]
 \begin{scriptsize}
  \begin{center}
\centering
\renewcommand{\arraystretch}{1.2}
\caption{Description of different source domains for the DIHARD III dataset.}
\label{Table:dtdesc}
\begin{tabular}{|c|c|c|c|c|l|}

 \hline
 \multirow{2}{*}{Domain} & Source & Speakers per & Total & Conversation & Description \\ 
 & Corpora & Recording & Speakers & Type &  \\ \hline \hline
 \multirow{2}{*}{Audiobooks (Ab)} &  \multirow{2}{*}{LIBRIVOX} &  \multirow{2}{*}{1} &  \multirow{2}{*}{12} &   \multirow{2}{*}{Read} & Amateur readers reading aloud passages from  public-domain\\  & & & & & English texts.  \\ \hline
 
\multirow{4}{*}{Broadcast} & \multirow{4}{*}{YOUTHPOINT} & \multirow{4}{*}{3-5} & \multirow{4}{*}{46} &  & Student-lead radio interviews conducted during the 1970s with \\ \multirow{4}{*}{Interview (Bi)} & & & & Radio & then-popular figures. The audio files are recorded in a studio on\\ & & & & Interview &  open reel tapes, and these were digitized and transcribed later\\ & & & & &  at LDC. \\
\hline

\multirow{3}{*}{Clinical (Cl)} & \multirow{3}{*}{ADOS} & \multirow{3}{*}{2} & \multirow{3}{*}{96} & \multirow{3}{*}{Interview} & Semi-structured interviews recorded with ceiling-mounted mic\\ & & & & &  to identify whether the 2-16 yrs children fit the clinical autism  \\ & & & & & diagnosis  \\ \hline

\multirow{4}{*}{Courtroom (Ct)} & \multirow{4}{*}{SCOTUS} & \multirow{4}{*}{5-10} & \multirow{4}{*}{83} & \multirow{4}{*}{Argument} & Oral arguments from the U.S. Supreme Court (2001 term). The\\ &  &  &  &  & table-mounted speaker-controlled mics' outputs were summed\\ &  &  &  &  & and recorded on a single-channel reel-to-reel \textcolor{black}{analog} tape recorder.\\ \hline

\multirow{1}{*}{CTS} & \multirow{1}{*}{FISHER} & \multirow{1}{*}{2} & \multirow{1}{*}{122} & \multirow{1}{*}{Telephonic} & Ten min conversations between two native English speakers. \\ \hline

\multirow{4}{*}{Map Task (Mt)} & \multirow{4}{*}{DCIEM } & \multirow{4}{*}{2} & \multirow{4}{*}{46} & \multirow{4}{*}{Formal} & Pairs of speakers sat opposite each other. One communicated a\\ &  &  &  &  & route marked on a map so that the other may precisely reproduce\\ &  &  &  &  & it on his map. Speech recorded on separate mics was mixed later.\\ \hline

\multirow{3}{*}{Meeting (Mg)} & \multirow{3}{*}{RT04/ROAR} & \multirow{3}{*}{3-10} & \multirow{3}{*}{75} &  & Highly interactive recordings consisting \textcolor{black}{of} large amounts of spon-\\ &  &  &  & {Formal} &-taneous speech with frequent interruptions and overlaps, each\\ &  &  &  &  & recorded with a centrally located distant mic with small gain.\\ \hline

\multirow{3}{*}{Restaurant (Rt)} & \multirow{3}{*}{CIR} & \multirow{3}{*}{5-8} & \multirow{3}{*}{86} & Interactive & Mix of the two channels of binaural mic recordings. Highly \\ &  &  &  & Informal & frequent overlap speech and interruptions along with speech\\ &  &  &  &  & from nearby tables and restaurant-typical background noise.\\ \hline

{Sociolinguistic} & \multirow{3}{*}{SLX/DASS} & \multirow{3}{*}{2-6} & \multirow{3}{*}{42} & \multirow{3}{*}{Interview} & An interviewer tried to elicit vernacular speech from an informant\\ (Field) (Sf) &  &  &  &  & during informal conversation. Most recordings are at home, while\\ &  &  &  &  & some are at public places (park or cafe). \\ \hline

{Sociolinguistic} & \multirow{2}{*}{MIXER6} & \multirow{2}{*}{2} & \multirow{2}{*}{32} & \multirow{2}{*}{Interview} & Controlled environment interviews recorded under quiet\\ (Lab) (Sl) &  &  &  &  & conditions. \\ \hline

\multirow{2}{*}{Web Video (Wv)} & {Video-sharing} & \multirow{2}{*}{1-9} & \multirow{2}{*}{127} & \multirow{2}{*}{Variable} & Collection of amateur videos on diverse topics and recording  \\  & sites &  &  &  & conditions in English and Mandarin. \\ \hline
\end{tabular}
  \end{center}
  \end{scriptsize}
\end{table}
\end{landscape}

\subsection{Signal-to-noise ratio (SNR) analysis}
\begin{figure}[t]
\centering
\includegraphics[width=3.5in]{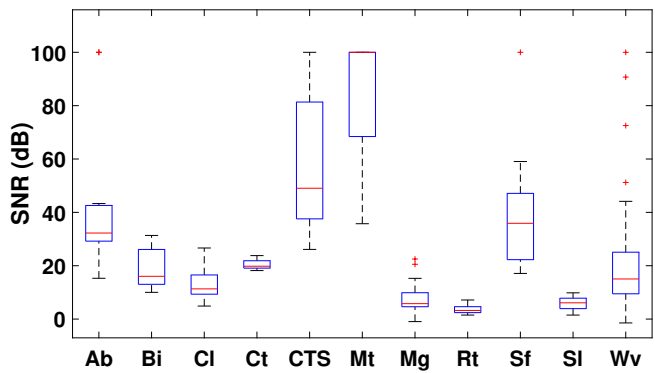}
\caption{SNR distribution of DIHARD III development data.}
\label{Fig:SNR}
\end{figure}

SNR is the most popular measure to quantify the presence of noise in signals. In SD, noisier audio samples are expected to be difficult to diarize. \emph{Waveform amplitude distribution analysis} (WADA) SNR estimation is a well-known technique for SNR estimation. It assumes that speech is totally independent of the background noise and clean speech always follows a gamma distribution with a fixed shaping parameter (between 0.4 and 0.5), whereas background noise follows Gaussian distribution \citep{kim2008robust}. The result of this estimation for the data in hand is shown in Fig.~\ref{Fig:SNR}. `CTS' and `Map task' show considerable variation in values and are on the higher side in the box-plot. This could be attributed to having only two speakers speaking on separate channels. `Meeting' and `Restaurant' have spontaneous speech from multiple speakers in noisy environment, recorded with one mic contributing to the low SNR. 

\begin{figure}[t]
\centering
\includegraphics[width=3.5in]{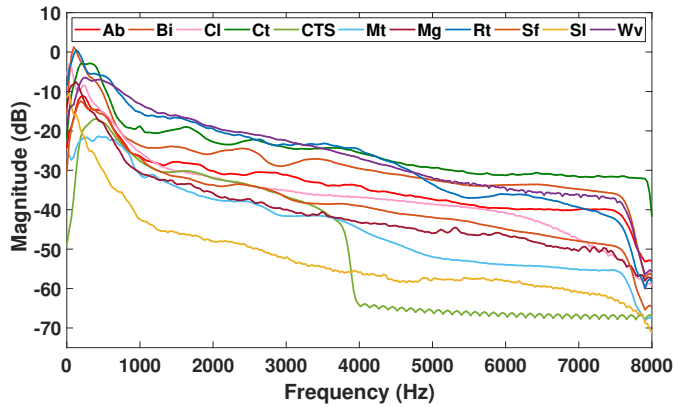}
\caption{Comparison of long-term average spectrum (LTAS) of speech files from each domain.}
\label{Fig:LTAS}
\end{figure}
\subsection{Long-term average spectrum (LTAS) analysis}
With LTAS, we are trying to gain knowledge of the spectral distribution of the signals over a period of time \citep{lofqvist1986long}. LTAS tends to average out segmental variations \citep{assmann2004perception}. This information is displayed in Fig.~\ref{Fig:LTAS} for DIHARD III development data in domain-wise manner. The fact that speech signals generally have no energy below 300~Hz but are low-frequency signals is discernible in all the plots. The peculiar nature of `CTS' after 4~kHz is because telephonic audio is sampled at 8~kHz. Controlled and quiet recording conditions like that of `Map task' and `Sociolinguistic (Lab)' show lower LTAS values. Note that the latter had low SNR too. On the other hand, multi-speaker and noisy environments of `Courtroom', `Restaurant', and `Web videos' exhibit higher magnitudes. 
\begin{figure}[t]
\centering
\includegraphics[width=3.5in]{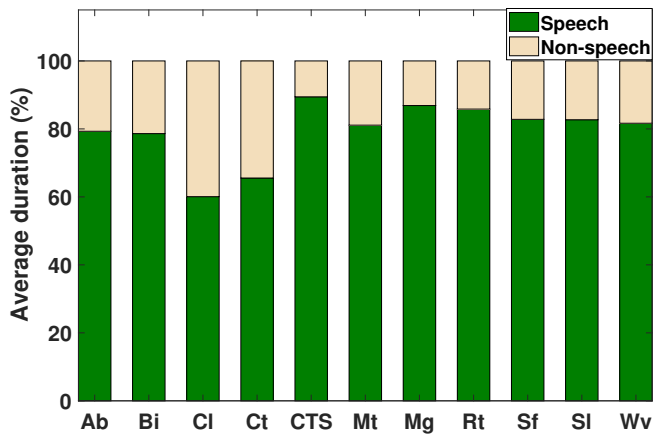}
\caption{Domain-wise percentage of Speech to non-speech average duration for DIHARD III development data.}
\label{Fig:hist_sp2nsp}
\end{figure}

\subsection{Speech-to-nonspeech ratio}
The final goal of this work is to assign time-stamps to spoken documents. So, the information regarding the amount of speech in the audio samples would be worthwhile. The speech-to-nonspeech ratio offers an acoustic representation of the language in daily conversations. It helps obtain additional information on the spectral energy distribution of a speech signal in a longer speech sample. As can be seen in Fig.~\ref{Fig:hist_sp2nsp}, around 20\% of the recorded signals from most of the domains are non-speech. The reason for exceptional behaviour of `Clinical' signals could be the unusual language use by children with autism, non-autistic language disorders, and ADHD. For `Courtroom' the high nonspeech value could be attributed to the participants' control on the table-top microphone and the overall noisy environment. 
\begin{figure}[t]
\centering
\includegraphics[width=3.5in]{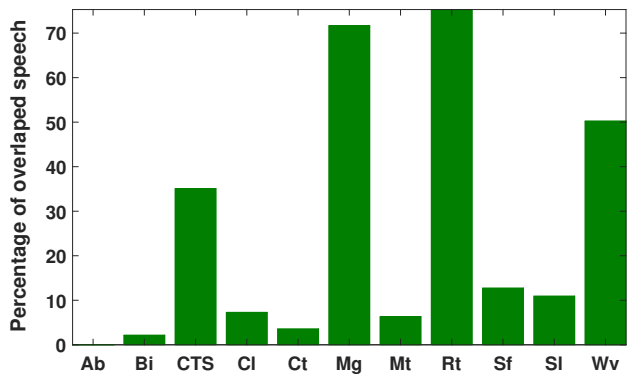}
\caption{Percentage of overlapped speech against non-overlapped speech averaged across all the samples of each domain for DIHARD III development data.}
\label{Fig:hist_olsp}
\end{figure}

\subsection{Percentage of overlapped speech}
The presence of overlapped speech in a spoken document is a serious factor affecting the performance of a diarization system. The SD systems are generally designed to assign one speaker label to a segment. Thus, when more than one speaker is present, the other speaker's or speakers' speech contributes to the \emph{missed detection rate} (MDR) component of the DER. A system having the capacity to assign multiple labels would also find it challenging to characterize particular speakers when their speech is superimposed. As can be seen in Fig.~\ref{Fig:hist_olsp}, overlapped speech amount varies with the domain. `Restaurant' has the highest value because many speakers are involved in informal conversation. In `Meeting', although the conversation is formal, the speakers can still interrupt each other resulting in high percentage of overlapped speech. In spite of multiple speakers having argument in `Courtroom', the reason for less overlapped speech is the same as that of high speech-to-nonspeech ratio.

\section{Baseline Diarization System}
\label{baseline}
A general SD system consists of an SAD, a segmentation, and a clustering module. In this work, we are concerned with only identifying the various domains of spoken documents and hence have only considered the Task 1 of DIHARD III \citep{ryant2020third} where the reference SAD was given. The diarization baseline provided with this challenge \citep{ryant2020thirdpaper}, which was based on one of the submissions of the predecessor challenge \citep{singh2019leap}, was used to benchmark our proposed SD system. 

The baseline system extracts 512-dimensional x-vectors using 30-dimensional MFCCs from 25~ms frames with overlap of 15~ms. Three second sliding windows are used for mean normalization. VoxCeleb 1 \citep{nagrani2017voxceleb} and VoxCeleb 2 \citep{chung2018voxceleb2} are combined and augmented with additive noise from MUSAN \citep{snyder2015musan} and reverberation from RIR datasets \citep{ko2017study} for training. Segments of duration 1.5~sec with shift of 0.25~sec are used for embeddings extraction during testing. Statistics estimated from DIHARD III Dev and Eval sets are used for centering and whitening of the embeddings, post which they are length normalized. 

Before scoring with Gaussian PLDA model \citep{prince2007probabilistic} trained from the x-vectors extracted from VoxCeleb, conversation-dependent PCA \citep{zhu2016online} preserving 30\% of the total variability is used for dimensionality reduction. Clustering is done using AHC with minimum DER for Dev set as the stopping criteria \citep{han2008strategies}. The output is then refined by using \emph{variational Bayes hidden Markov model} (VB-HMM) re-segmentation \citep{diez2019analysis} which is initialized separately for each recording and run for one iteration. For this, 24-dimensional MFCCs are used without mean or variance normalization nor delta coefficients, extracted from 15~ms frames with 5~ms overlap. A 1024-diagonal covariance components GMM-\emph{universal background model} (UBM) and 400 eigenvoices TV matrix, both trained with the x-vector extractor data, are used. The zeroth-order statistics are boosted before VB-HMM likelihood computation for posterior scaling to reduce frequent speaker transitions \citep{singh2019leap}.  

\section{Domain classification}
\label{sec:classify}
\begin{figure*}[t]
\centering
\includegraphics[width=0.85\linewidth]{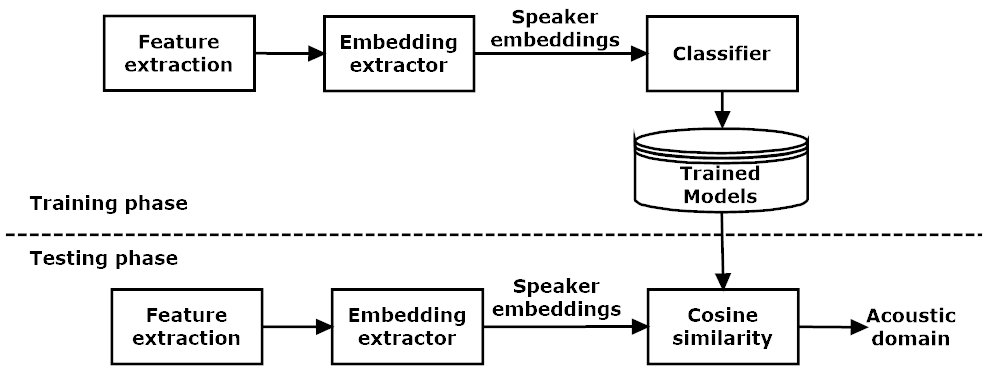}
\caption{Block diagram of the proposed acoustic domain identification system.}
\label{Fig:BlockADI}
\end{figure*}

The framework of proposed ADI system is shown in Fig.~\ref{Fig:BlockADI}. It is based on the speaker embeddings as recording-level feature and nearest neighbor classifier. Though the speaker embeddings are principally developed for speaker characterization, they also capture information related to acoustic scene~\citep{zeinali2018convolutional}, recording session~\citep{Probing2019}, and channel~\citep{Wang2017}. In this work, we studied two frequently used speaker embeddings: discriminatively trained x-vectors and generative i-vectors. In an earlier study, these speaker embeddings were investigated for the second DIHARD dataset on related tasks~\citep{sahidullah2019speed, fennir2020acoustic}. Owing to the similarity of ADI with ASC, we additionally investigated L3-net embeddings, which were a part of the baseline system provided with the ASC task of the DCASE 2020 challenge \citep{Heittola2020}.

\subsection{i-vector embeddings}
Speaker diarization involves splitting multi-speaker spoken documents according to speaker changes and segment clustering. Since it closely resembles speaker recognition, the latter’s algorithms are applied to the former with good results. i-vectors are one such example~\citep{dehak2010front}. Speaker recognition systems work in two steps, enrolment and recognition. During the speaker enrolment process, a UBM~\citep{reynolds2000speaker} is generated using the data collected from non-target utterances. A target model is generated by adapting the UBM according to the target data. Whenever any test utterance is given as input to the system, features are extracted from it, and pattern matching algorithms are applied using one or more kinds of target models. i-vectors is designed to improve the joint factor analysis~\citep{kenny2007joint} by combining the inter and intra-domain variability and modeling it in the same low dimensional total variability space. This space $\mathbf{T}$ accounts for both speaker and channel variability.

The speaker and channel dependent supervector $\mathbf{M}$ is given by

\begin{equation} \label{eqn1}
\mathbf{M} = \mathbf{m}+\mathbf{T}\mathbf{x}
\end{equation}

\noindent If $C$ is the number of components in UBM and $F$ is the dimension of acoustic feature vectors then for a given utterance, we concatenate the $F$-dimensional GMM mean vectors to get the speaker and channel independent UBM supervectors $\mathbf{m}$ of dimension $CF$. The normally distributed random vectors $\mathbf{x}$ are known as identity vectors or i-vectors. It is assumed that $\mathbf{M}$ is distributed normally with $\mathbf{T}\mathbf{T}^{\top}$ as its covariance matrix and $\mathbf{m}$ as its mean vector. For training of the matrix $\mathbf{T}$, it is assumed that the utterances of the same speaker are produced by different speakers.

\subsection{x-vector embeddings}
x-vectors represent a recent but popular approach from speaker recognition research~\citep{snyder2018x, BAI202165}. Here, long-term speaker characteristics are captured in a feed-forward deep neural network by a temporal pooling layer that aggregates over the input speech \citep{snyder2017deep}. The network has layers that operate on speech frames, a statistics pooling layer that aggregates over the frame-level representations, additional layers that operate at the segment-level, and a softmax output layer at the end. The nonlinearities are \emph{rectified linear units} (ReLUs). 

Suppose an input segment has $K$ frames. The first five layers operate on speech frames, with a small temporal context centered at the current frame. The statistics pooling layer aggregates all $K$ frame-level outputs from fifth frame-layer and computes its mean and standard deviation. This process aggregates information across the time dimension so that subsequent layers operate on the entire segment.The mean and standard deviation are concatenated together and propagated through segment-level layers, and finally, the softmax output layer.

The goal of training the network is to produce embeddings that generalize well to speakers that have not been seen in the training data. The embeddings should capture speaker characteristics over the entire utterance, rather than at the frame-level. After training, the embeddings extracted from the affine component of first segment-layer are referred to as x-vectors. The pre-softmax affine layer is not considered because of its large size and dependence on the number of speakers.

\subsection{L3-net embeddings}
The look, listen and learn (L3)-network embeddings were proposed for learning \emph{audio-visual correspondence} (AVC) in videos in a self-supervised manner \citep{arandjelovic2017look}. The goal of AVC learning was to \emph{learn} audio and visual semantic information simultaneously by simply \emph{watching} and \emph{listening} to many unlabelled videos. Without needing annotated data and a relatively simple convolutional architecture, L3-net successfully produced powerful embeddings that led to state-of-the-art sound classification performance. Thus, it can be applicable to a wide variety of machine learning tasks with scarce labeled data. However, in \citep{arandjelovic2017look}, non-trivial parameter choices that may affect the performance and computation cost of the model were not elaborated. This issue was addressed in \citep{cramer2019look} in the context of sound event classification. The authors worked with three publicly available environmental audio datasets and found that mel-spectrogram gave better audio representation than linear spectrogram, which was used in \citep{arandjelovic2017look}, Their results also indicated that for a downstream classification task, using the audio that makes the embeddings most discriminative, independently of the downstream domain, is more important. Regarding the training data, it was suggested that at least 40M samples should be used to train the L3-Net embedding. 

\par
We first computed the average of the embeddings for each domain and stored it as trained models. We calculated the cosine similarity during testing and assigned the class label with the highest similarity.


\section{Experimental setup}
\label{sec:expt}
The ADI experiments were performed on the development set of 254 speech utterances from 11 domains. We randomly selected 200 utterances for training and used the remaining 54 for test. For similarity measurement between training and test data, \emph{cosine similarity} was employed. The experiments were repeated 1000 times. 

To extract utterance-level embeddings for the ADI task, we used pre-trained x-vector and i-vector models trained on VoxCeleb audio-data\footnote{\url{https://kaldi-asr.org/models/m7}}. Pre-trained versions of the L3-Net variants studied in \citep{cramer2019look} are made freely available online by the name OpenL3\footnote{\url{https://github.com/marl/openl3}}. Open L3 embeddings were used as feature representation for the baseline system provided with the DCASE 2020's ASC task \citep{Heittola2020}. The system had a one-second analysis window, with a 100~ms hop, 256 mel filters for input representation, content type music, and a 512-dimensional embedding. We used the same for our ADI task. 

\begin{figure*}[t]
\centering
\includegraphics[trim={1cm 0 0 0},clip,width=0.95\linewidth]{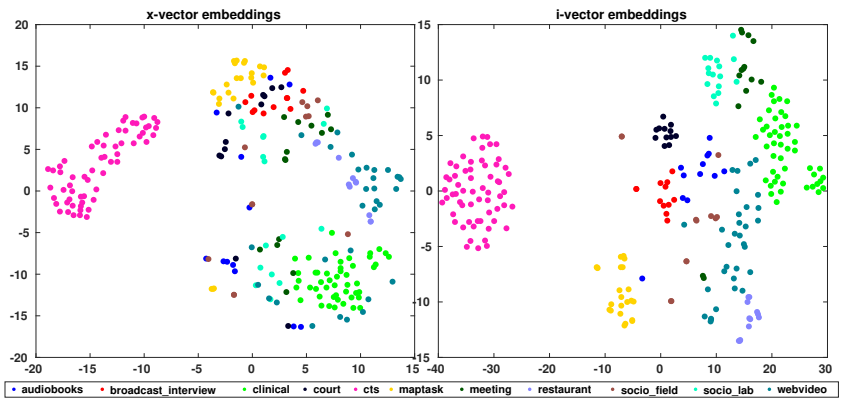}
\caption{\textcolor{black}{Domain-wise scatter plot of x-vector and i-vector features.}}
\label{Fig:tSNE}
\end{figure*}

\begin{figure}[t]
\centering
\includegraphics[trim={0 1.1cm 0 0},clip,width=4.7in]{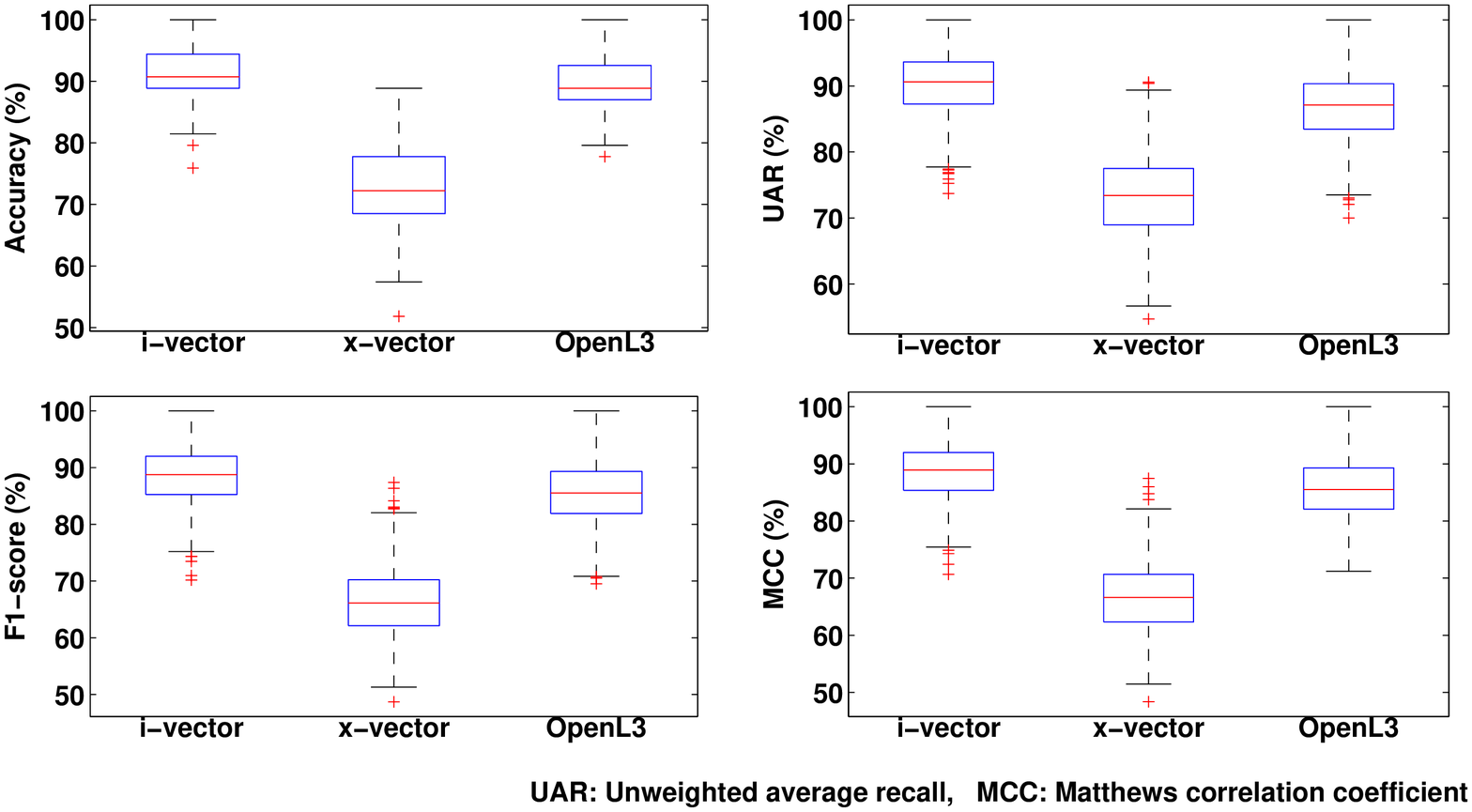}
\caption{Acoustic domain identification performance in terms of accuracy, F1-score, unweighted average recall (UAR),  Matthews correlation coefficient (MCC) using i-vector, x-vector, and OpenL3 embeddings.}
\label{Fig:ADIaccuracy}
\end{figure}

\textcolor{black}{We performed domain-dependent speaker diarization on the evaluation set of DIHARD III data by first predicting the domain for every utterance using the ADI system and grouping the utterances according to the predicted domains. i-vector embeddings manifest superior domain-clustering abilities (refer Fig.~\ref{Fig:tSNE} and Fig.~\ref{Fig:ADIaccuracy} in Section~\ref{sec:result_adi}), so we resort to them for ADI. We used domain-specific thresholds for speaker clustering in the diarization pipeline. Towards this, we computed the domain-specific speaker diarization performance for different thresholds of the development set. We selected the threshold value that shows the best SD performance for a specific subset.}

For diarization, our experimental setup is based on the baseline system created by the organizers~\citep{ryant2020thirdpaper}. We have used the toolkit\footnote{\url{https://github.com/dihardchallenge/dihard3_baseline}} with the same frame-level acoustic features, embedding extractor, scoring method, etc. The system is trained with all 254 sentences for identifying the domains of evaluation data samples.

The primary evaluation metric of the challenge is \emph{diarization error rate} (DER), which is the sum of all the errors associated with the diarization task. \emph{Jaccard error rate} (JER) is the secondary evaluation metric, and it is based on the Jaccard similarity index. The details of the challenge with rules are available in the challenge evaluation plan~\citep{ryant2020third}.

\section{Results}
\label{sec:result}

\subsection{ADI experiments}
\label{sec:result_adi}
First, we present in Fig.~\ref{Fig:tSNE} the domain-wise clustering ability of the two speaker embeddings: i-vectors and x-vectors. The scatter-plots show that `CTS' samples are well-separated from the rest of the data because this subset consists of narrowband telephone speech, whereas the others are from wideband speech. We also observe that i-vector embeddings have done better for the other domains than the x-vector embeddings. However, for `Sociolinguistic (Field)', the clusters are not well represented by both the embeddings, likely due to the within-class variability in the subset (refer Table~\ref{Table:dtdesc}, Row 9). 

We carried out ADI with the three embeddings described in Section~\ref{sec:classify}. Figure~\ref{Fig:ADIaccuracy} shows the corresponding performance comparison. Besides the commonly used classification accuracy, we also reported \emph{unweighted average recall} (UAR), \emph{F1-score}, and \emph{Mathews correlation coefficient} (MCC), because of the unbalanced nature of the data. The figure shows that the i-vector and OpenL3 systems are substantially better than the x-vector system for ADI on the DIHARD III dataset. With the present experimental setup, i-vectors surpass the other two embeddings on all four metrics. For instance, the average domain classification accuracy over 1000 repetitions was 91.11\%, 72.98\%, and 89.64\% for i-vector, x-vector, and OpenL3 systems, respectively. The poor performance of x-vectors could be because they are originally designed to capture better speaker characteristics over the entire utterance. However, here we are trying to classify domains where one utterance may have multiple speakers, and the same domain data may not have common speakers. 

The ADI results over evaluation data of the DIHARD III challenge for the three embeddings are displayed in Table~\ref{Table:adi_eval}. Here, we notice that the four metrics follow the same pattern as in Fig.~\ref{Fig:ADIaccuracy}, but the difference in the performance of the embeddings is more prominently visible. Based on these values, we chose i-vectors for ADI and proceeded to the speaker diarization stage.

\begin{table}[]
\caption{Performance of ADI system on DIHARD III evaluation for i-vector, x-vector, and OpenL3 based systems.}
\label{Table:adi_eval}
\begin{center}
\begin{tabular}{|c|c|c|c|c|}
\hline
\hline
        & Accuracy (\%) & F1-Score (\%) & UAR (\%) & MCC (\%) \\ \hline
        \hline
i-vector & \textbf{86.49}         & \textbf{80.26}         & \textbf{82.76}    & \textbf{80.24}             \\ \hline
x-vector & 69.88                  & 56.61                  & 60.14             & 55.85             \\ \hline
OpenL3   & 75.29                  & 72.84                  & 73.43             & 71.66             \\ \hline
\end{tabular}
\end{center}
\end{table}

\begin{table}[!t]
  \caption{The impact of domain-dependent processing on speaker diarization performance (DER in \%/JER in \%) for different acoustic domains of the development set of the third DIHARD challenge. (P1: Domain-dependent threshold and PLDA adaptation, P2: Domain-dependent threshold but PLDA adaptation with full data.)}
  
  \begin{scriptsize}
  \begin{center}
  \label{Table:ResultsSubset}
  \centering
  \renewcommand{\arraystretch}{1.2}
\begin{tabular}{|c|c|c|c|c|c|}

\hline
  &                          & \multicolumn{2}{c|}{First Pass} & \multicolumn{2}{c|}{Re-segmentation with VB-HMM} \\ \cline{3-6} 
\multirow{-2}{*}{Domain}  & \multirow{-2}{*}{Method} & Full           & Core           & Full                    & Core                   \\ \hline \hline
 & Baseline                 & 4.95 / 4.81    & 4.95 / 4.81    & 4.55 / 4.45             & 4.55 / 4.45            \\  
 & P1                       & \textbf{0.00 / 0.00}    & \textbf{0.00 / 0.00}    & \textbf{0.40 / 0.40}             & \textbf{0.40 / 0.40}            \\  
\multirow{-3}{*}{Audiobooks}                                                    & P2                       & \textbf{0.00 / 0.00}    & \textbf{0.00 / 0.00}    & \textbf{0.40 / 0.40}             & \textbf{0.40 / 0.40}            \\ \hline
 & Baseline                 & 3.75 / 25.62   & 3.75 / 25.62   & 3.56 / 23.83            & 3.56 / 23.83           \\  
 & P1                       & 6.47 / 39.38   & 6.47 / 39.38   & 6.10 / 38.98            & 6.10 / 38.98           \\  
\multirow{-3}{*}{\begin{tabular}[c]{@{}c@{}}Broadcast\\ interview\end{tabular}} & P2                       & \textbf{3.51 / 24.05}   & \textbf{3.51 / 24.05}   & \textbf{3.29 / 22.27}            & \textbf{3.29 / 22.27}           \\ \hline
 & Baseline                 & 17.55 / 28.88  & 16.08 / 26.38  & 17.69 / 28.46           & 16.72 / 27.21          \\  
 & P1                       & 20.06 / 29.92  & 18.88 / 28.11  & 16.71 / 25.50           & 15.21 / 23.66          \\  
\multirow{-3}{*}{Clinical}                                                      & P2                       & \textbf{15.81 / 23.69}  & \textbf{14.61 / 22.37}  & \textbf{14.69 / 22.07}           & \textbf{13.78 / 21.59}          \\ \hline
 & Baseline                 & 10.81 / 38.75  & 10.81 / 38.75  & 10.17 / 37.63           & 10.17 / 37.63          \\  
 & P1                       & 12.19 / 43.99  & 12.19 / 43.99  & 9.03 / 37.97            & 9.03 / 37.97           \\ 
\multirow{-3}{*}{Courtroom}                                                     & P2                       & \textbf{5.82 / 23.91}   & \textbf{5.82 / 23.91}   & \textbf{4.77 / 22.22}            & \textbf{4.77 / 22.22}           \\ \hline
 & Baseline                 & 22.25 / 27.85  & 23.48 / 28.41  & 18.38 / 22.32           & 18.84 / 22.37          \\  
 & P1                       & 20.22 / 27.73  & 20.28 / 27.18  & 17.31 / 22.31           & 16.89 / 21.07          \\  
\multirow{-3}{*}{CTS}                                                           & P2                       & \textbf{19.23 / 25.93}  & \textbf{19.81 / 25.96}  & \textbf{16.55 / 20.94}           & \textbf{17.10 / 21.20}          \\ \hline
   & Baseline                 & 9.92 / 18.90   & 9.92 / 18.90   & 8.75 / 16.66            & 8.75 / 16.66           \\  
  & P1                       & 12.19 / 25.75  & 12.19 / 25.75  & 9.23 / 20.25            & 9.23 / 20.25           \\  
\multirow{-3}{*}{Map Task}& P2                       & \textbf{9.21 / 18.20}   & \textbf{9.21 / 18.20}   & \textbf{8.50 / 16.50}            & \textbf{8.50 / 16.50}           \\ \hline
  & Baseline                 & \textbf{29.65} / 51.57  & \textbf{29.65} / 51.57  & 28.80 / 51.01           & 28.80 / 51.01          \\  
  & P1                       & 31.31 / 52.71  & 31.31 / 52.71  & 29.35 / 50.85           & 29.35 / 50.85          \\  
\multirow{-3}{*}{Meeting} & P2                       & 29.80 / \textbf{50.72}  & 29.80 / \textbf{50.72}  & \textbf{28.62 / 50.09}           & \textbf{28.62 / 50.09}          \\ \hline
  & Baseline                 & 46.99 / 73.80  & 46.99 / 73.80  & 46.60 / 74.83           & 46.60 / 74.83          \\  
  & P1                       & 48.68 / 74.11  & 48.68 / 74.11  & 46.40 / 73.83           & 46.40 / 73.83          \\  
\multirow{-3}{*}{Restaurant}                                                                            & P2                       & \textbf{45.34 / 70.49}  & \textbf{45.34 / 70.49}  & \textbf{45.34 / 72.98}           & \textbf{45.34 / 72.98}          \\ \hline
  & Baseline                 & 17.67 / \textbf{49.48}  & 17.67 / \textbf{49.48}  & 16.24 / \textbf{47.68}           & 16.24 / \textbf{47.68}          \\  
  & P1                       & 23.06 / 60.61  & 23.06 / 60.61  & 21.53 / 58.31           & 21.53 / 58.31          \\  
\multirow{-3}{*}{\begin{tabular}[c]{@{}c@{}}Sociolingusitic\\ (Field)\end{tabular}}                     & P2                       & \textbf{17.46} / 51.19  & \textbf{17.49} / 51.19  & \textbf{16.16} / 49.16           & \textbf{16.16} / 49.16          \\ \hline
   & Baseline                 & 11.97 / 17.65  & 11.97 / 17.65  & 10.33 / 14.94           & 10.33 / 14.94          \\  
  & P1                       & 13.60 / 23.84  & 13.60 / 23.84  & 11.41 / 18.70           & 11.41 / 18.70          \\  
\multirow{-3}{*}{\begin{tabular}[c]{@{}c@{}}Sociolinguistic\\ (Lab)\end{tabular}}                       & P2                       & \textbf{10.38 / 15.92}  & \textbf{10.38 / 15.92}  & \textbf{9.21 / 13.87}            & \textbf{9.21 / 13.87}           \\ \hline
  & Baseline                 & 40.99 / 76.97  & 40.99 / 76.97  & \textbf{40.06 / 76.77}          & \textbf{40.06 / 76.77}          \\  
  & P1                       & 41.60 / 79.93  & 41.60 / 79.93  & 40.97 / 79.66           & 40.97 / 79.66          \\  
\multirow{-3}{*}{Web Video}                                                                             & P2                       & \textbf{40.14 / 79.29}  & \textbf{40.14 / 79.29}  & 40.25 / 79.38           & 40.25 / 79.38          \\ \hline
\end{tabular}
 
  \end{center}
  \end{scriptsize}
\end{table}

\subsection{Speaker diarization experiments: Development set}
\label{sec:resultsanalysis}

In our first SD experiment with the baseline system, we evaluated the development set of the DIHARD III dataset and analyzed the diarization performance for each domain separately. The top row for each domain listed in Table~\ref{Table:ResultsSubset} shows the domain-wise baseline results in terms of DER and JER. As expected, we notice that the SD performance varied with domains. The DER values range from less than $4\%$ for `Broadcast Interview' domain to greater than $46\%$ corresponding to 'Restaurant' domain.

Next, we performed the diarization experiments with domain-dependent processing. We selected a domain-dependent threshold for speaker clustering. In addition, for PLDA adaptation during the scoring process, we considered audio data from specific domains only. This system is referred to as \textbf{P1} in Table~\ref{Table:ResultsSubset}. To our surprise, this degraded the performance compared to the baseline for most of the domains. We hypothesize that the limited speaker variability in domain-wise audio data (refer to Table~\ref{Table:dtdesc}) could be a factor as we used this data as in-domain target data for PLDA adaptation. The PLDA adaption with domain-specific data was helpful for `Audiobooks' and `CTS' because the two domains are substantially different from the others. For example, `Audiobooks' consists of high-quality recordings with a single speaker, and `CTS' consists of narrow-band telephone speech with two speakers.

In the next set of SD experiments, we performed PLDA adaptation with audio data from all the eleven subsets while applying domain-specific thresholds for speaker clustering. We refer to this method as \textbf{P2} in Table~\ref{Table:ResultsSubset}. The noticeable improvement in `Courtroom' audio diarization could be because of the following reason. As pointed out earlier in Section~\ref{sec:adi}, despite having multiple speakers in argument, `Courtroom' has less overlapped speech and a low value of the speech-to-nonspeech ratio. Another interesting characteristic of this domain's data is its high spectral energy, even at high frequencies, along with only a slight variation in the SNR. The minor improvement in the DER and worsened JER for `Sociolinguistic (Field)' reiterates the poor clustering of this domain's samples (refer Fig.~\ref{Fig:tSNE}). A similar argument applies to the trends of results in `Meeting' and `Web videos.'

After the first pass for the `Full' data condition, we observe that `Audiobooks,' `Courtroom,' and `CTS' exhibit a major relative improvement of 100\%, 46.16\%, and 13.57\% in DER, respectively, to the baseline system. However, following re-segmentation (second pass), the highest improvement in DER is exhibited by `Audiobooks': 91.21\%, `Clinical': 16.96\%, and `Courtroom': 53.09\%, respectively. These domains have less speaker overlap and lower speech-to-nonspeech ratios except the `CTS' domain. On the other hand, a poor relative improvement of -0.50\%, 1.19\%, and 2.07\% in DER is registered for `Meeting,' `Sociolinguistic (Field),' and `Web video' respectively after the first pass of diarization. The same domains show low relative DER improvements of 0.63\%, 0.49\%, and -0.47\%, respectively, after the second pass of diarization. This can be explained by the fact that these domains have considerable overlapping speech and poor SNR. For most domains, the relative improvement is higher after the first pass than after the second because the clustering was done in the first pass at an ideal threshold, leaving little room for improvement during re-segmentation.

Our detailed analysis of the development set results indicates that the \textbf{P2} method shows reduced DER/JER values for eight out of eleven domains. This confirms our hypothesis that domain-dependent threshold for speaker clustering helps. We subsequently used \textbf{P2} method for diarization of the evaluation set.

\begin{table}[t]
  \caption{Results showing the speaker diarization performance using baseline and proposed method on development and evaluation set of third DIHARD challenge. For this phase, we consider the second pass with re-segmentation as it consistently gives superior SD performance over the first pass.}
  \begin{footnotesize}
  \begin{center}
  \label{Table:Dev}
  \centering
  \renewcommand{\arraystretch}{1.2}
  \begin{tabular}{|c|c|c|c|c|c|}
    \hline
    \multirow{2}{*}{Set} & \multirow{2}{*}{Method} & \multicolumn{2}{|c|}{Full} & \multicolumn{2}{|c|}{Core}\\
     \cline{3-6}
     & & DER & JER & DER & JER \\
    \hline
    \hline
    \multirow{2}{*}{Development}&Baseline	& 19.59	&43.01	&20.17&	47.28\\
    &Proposed	& 17.97	&40.33	&18.73	&44.77\\
    \hline
        \multirow{2}{*}{Evaluation}&Baseline	& 19.19	&43.28&	20.39&	48.61\\
    &Proposed	& 17.56	&38.60&	19.23&	43.74\\
    \hline
  \end{tabular}
  \vspace{-0.5cm}
  \end{center}
  \end{footnotesize}
\end{table}

\subsection{Speaker diarization experiments: Evaluation set}

We performed experiments on the evaluation set by predicting the domain for each of its recordings which is followed by selecting the corresponding clustering threshold. For domain prediction, we used the i-vector-based embeddings as features and the nearest neighbor classifier as this gave promising results for ADI (See details in Section~\ref{sec:result_adi}).
 
Table~\ref{Table:Dev} summarizes the results for the entire development and evaluation sets. We observe substantial overall improvements in development and evaluation data. For the `Full' condition of the evaluation set consisting of all the audio files, we obtained a relative improvement of 8.49\% and 10.81\% in terms of DER and JER, respectively. Whereas for the `Core' condition with a balanced amount of audio data per domain, we achieved a relative improvement of 5.69\% and 10.02\%. This reflects the results of the detailed analysis of individual domains in Section~\ref{sec:resultsanalysis} where we established that the relative improvements with the proposed method varied for different domains.

\section{Conclusion}
\label{sec:conc}
This work advances the state-of-the-art DNN-based speaker diarization performance with the help of domain-specific processing. We applied speech embeddings computed from the entire audio recording for the supervised audio domain identification task. After experimenting with three different embeddings, we employed i-vector embeddings and the nearest neighbor classifier for the ADI task. We integrated the ADI system with the x-vector based state-of-the-art diarization system and performed the experiments on the DIHARD III challenge dataset consisting of multiple domains. We obtained up to 8\% relative improvement on the `Full' condition of the evaluation set. The proposed method substantially improved the diarization performance on most of the subsets of the DIHARD III dataset.

Our work involved the application of domain-specific thresholds for speaker clustering. This study can be extended by incorporating other domain-specific operations such as speech enhancement and feature engineering. We used a basic \emph{time delay neural network} (TDNN) based x-vector embedding for audio segment representation. The future work may also include investigation of more advanced embeddings (e.g., \emph{emphasized channel attention, propagation, and aggregation in TDNN} or ECAPA-TDNN), which have shown promising results in related tasks.

The present study is limited to a supervised domain classification problem where the training phase requires class-annotated data. However, in practice, the audio data may come from myriad unknown sources, and our domain classification method may not be applicable in its current form. Nevertheless, an investigation of unsupervised domain clustering to exploit the advantage of domain-dependent processing in such a scenario can be carried out.

\section*{Acknowledgements}

\textcolor{black}{We thank the anonymous reviewers for their careful reading of our manuscript and their insightful comments and suggestions.} This work is funded by the Jagadish Chandra Bose Centre of Advanced Technology (JCBCAT), Govt. of India. Experiments presented in this paper were partially carried out using the Grid'5000 testbed, supported by a scientific interest group hosted by Inria and including CNRS, RENATER and several Universities as well as other organizations (see \url{https://www.grid5000.fr}).



\bibliographystyle{aps-nameyear}      
\bibliography{mybib.bib}      

\end{document}